\documentclass[runningheads]{llncs}
\usepackage[T1]{fontenc}
\usepackage{graphicx,verbatim}
\usepackage{kotex}

\usepackage{graphicx}
\usepackage{amsmath}
\usepackage{amssymb}
\usepackage{booktabs}
\usepackage{multirow}
\usepackage{paralist}
\usepackage{array}
\usepackage[normalem]{ulem}
\usepackage[accsupp]{axessibility}
\usepackage{soul}
\usepackage{url}
\usepackage{color, colortbl}
\usepackage{relsize}
\usepackage{subcaption}
\usepackage{xspace}
\usepackage[pagebackref,breaklinks,colorlinks=false,hidelinks,bookmarks=false]{hyperref}

\usepackage[capitalize]{cleveref}
\crefname{section}{Sec.}{Secs.}
\Crefname{section}{Section}{Sections}
\Crefname{table}{Table}{Tables}
\crefname{table}{Tab.}{Tabs.}

\usepackage[usenames,dvipsnames]{xcolor}

\newcommand{\blockcomment}[1]{}

\newcommand{\FIXME}[1]{\textcolor{red}{#1}\xspace}

\newcommand{\CITEME}[1]{\FIXME{\cite{CITEME}}\xspace}
\newcommand{\OURS}{DiGDA\xspace}

\makeatother

\begin{document}
\title{
    Diffusion-Based User-Guided Data Augmentation
    for Coronary Stenosis Detection
}
%
\author{Sumin Seo\inst{1}\orcidID{0000-0001-8703-0322}\thanks{Corresponding author: sumin.seo@medipixel.io} \and
In Kyu Lee\inst{1,2}\orcidID{0000-0001-5554-808X} \and\\
Hyun-Woo Kim\inst{1}\orcidID{0009-0003-2740-0397} \and
Jaesik Min\inst{1}\orcidID{0000-0002-0007-0637} \and
Chung-Hwan Jung\inst{1}
}
\authorrunning{S. Seo et al.}
%
\institute{Medipixel, Inc., Republic of Korea\\
\email{{\{sumin.seo,hyunwoo.kim,jaesik.min,danny.jung\}}@medipixel.io}\\ \and
University of California San Diego, USA\\
\email{kl002@ucsd.edu}}


\maketitle              

\begin{abstract}

Coronary stenosis is a major risk factor for ischemic heart events leading to increased mortality, and medical treatments for this condition require meticulous, labor-intensive analysis. Coronary angiography provides critical visual cues for assessing stenosis, supporting clinicians in making informed decisions for diagnosis and treatment.
Recent advances in deep learning have shown great potential for automated localization and severity measurement of stenosis. In real-world scenarios, however, the success of these competent approaches is often hindered by challenges such as limited labeled data and class imbalance. In this study, we propose a novel data augmentation approach that uses an inpainting method based on a diffusion model to generate realistic lesions, allowing user-guided control of severity. 
Extensive evaluation on lesion detection and severity classification across various synthetic dataset sizes shows superior performance of our method on both a large-scale in-house dataset and a public coronary angiography dataset.
Furthermore, our approach maintains high detection and classification performance even when trained with limited data, highlighting its clinical importance in improving the assessment of severity of stenosis and optimizing data utilization for more reliable decision support.

\keywords{Stenosis detection \and Coronary angiography \and Diffusion models.}

\end{abstract}

\section{Introduction}

\begin{figure}[ht]
\centering
    \includegraphics[width=\linewidth]{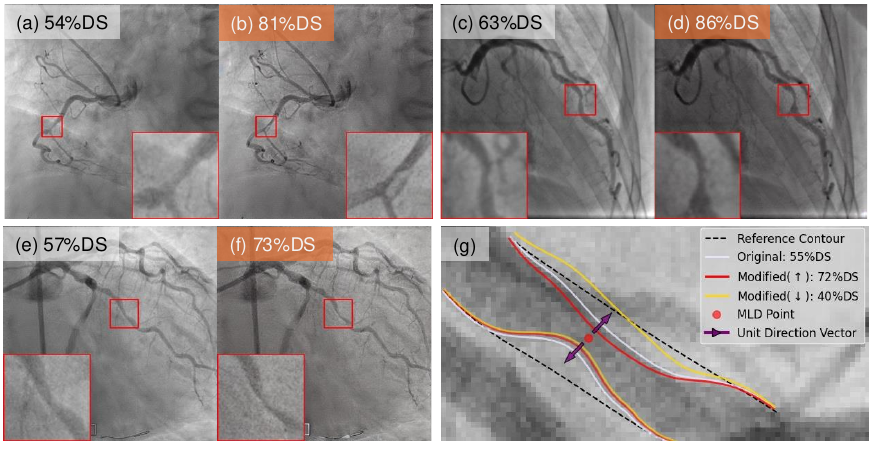}
    \caption{
        \textbf{Examples of coronary angiograms generated with various levels of stenosis.} (a,c,e) are the original angiograms and (b,d,f) are the angiograms generated with specific \%DS values applied to the vessel segmentation maps. (g) describes how the automatic algorithm generates a mask when numeric value of \%DS is given by user. Minimum lumen diameter (MLD) point is marked as a red circle in the image and represents the center of the bounding boxes.
    }
    \label{fig:qualitative}
\end{figure}

Coronary artery disease (CAD), a leading cause of global mortality~\cite{ahmad2024mortality,di2024heart}, is characterized by narrowed artery often caused by buildup of plaque, reducing blood flow to the heart. Coronary stenosis, a key manifestation of CAD, necessitates accurate assessment of the severity of stenosis in coronary angiography (CAG) for making informed clinical decisions and planning interventions. 
Among the various metrics used to quantify stenosis, the percentage diameter stenosis (\%DS) plays a central role: a \%DS exceeding 50\% is typically considered clinically significant, while lesions above 70\% are generally classified as severe~\cite{writing20222021}.
Different severity levels often lead to diagnostic decision or therapeutic interventions.

Deep learning-based approaches to automatic stenosis detection and severity classification using angiograms have shown promising results~\cite{danilov2021real,jimenez2024cadica,lin2022deep,moon2021automatic} in recent years. 
However, practical adaptation often encounters several critical challenges: data scarcity, class imbalance, and labeling cost. 
While large-scale medical datasets are publicly available for other imaging modalities, such as chest X-ray, CT, and MRI~\cite{li2020lungpetctdx,pham2023pedicxr,sunoqrot2022artificial}, the largest publicly available angiogram dataset contains only around 4,000 images from at most 42 patients~\cite{jimenez2024cadica}. 
In addition, typical CAG datasets indicate that a majority of the cases are moderate (non-severe) stenosis rather than severe ones~\cite{popov2024arcade}.
This distribution biases neural networks toward the majority class and impedes their ability to detect severe cases~\cite{lin2023stenunet}.
Another major limitation is the high labeling cost associated with stenosis severity annotation. Manual severity grading is labor-intensive and subject to significant variation across observers~\cite{shivaie2024interobserver}, which introduces labeling ambiguity. This ambiguity necessitates additional expert review and consensus-building, further increasing the overall cost of obtaining reliable annotations for robust model training~\cite{sylolypavan2023impact}.

One promising strategy to address these issues is to augment the data by synthesizing images. Previous studies have employed segmentation masks to synthesize anatomically realistic medical images~\cite{Cob_Improved_MICCAI2024,konz2024anatomically}. However, unlike classification problems in other medical imaging domains, lesions in angiograms can occur at any location within the vessels, making it impractical to rely solely on segmentation masks for lesion detection. 
Since tracking all vessel branches is infeasible, using only main branch segmentation masks for image generation can lead to unintended lesion artifacts in non-target regions. If these unintended lesions are not explicitly labeled, they introduce noise into the dataset, potentially degrading model performance~\cite{fang2024data}.
In contrast, our synthesis method aims at specific regions of interest (ROIs), allowing user-defined modifications to the lesion areas. 
By restricting changes to the ROIs, our approach synthesizes realistic variations in stenosis severity without introducing unintended artifacts in unrelated regions. 
This targeted augmentation, which is suitable for lesion detection training data, better reflects the diverse manifestations of severe lesions without the need for additional labeling cost. 

To our knowledge, this work is the first to integrate an inpainting strategy with a generative model that synthesizes angiograms for lesion detection. The generated images preserve realistic vascular structure, as illustrated in Fig.~\ref{fig:qualitative}.
Utilizing user-defined masks to make a balanced class distribution enhances model performance compared to training without them.
Extensive experimental results consistently demonstrate our pipeline outperforms the model without any synthetic data under data-scarce conditions and across various model sizes.
\begin{figure}[t]
\centering
    \includegraphics[width=\linewidth]{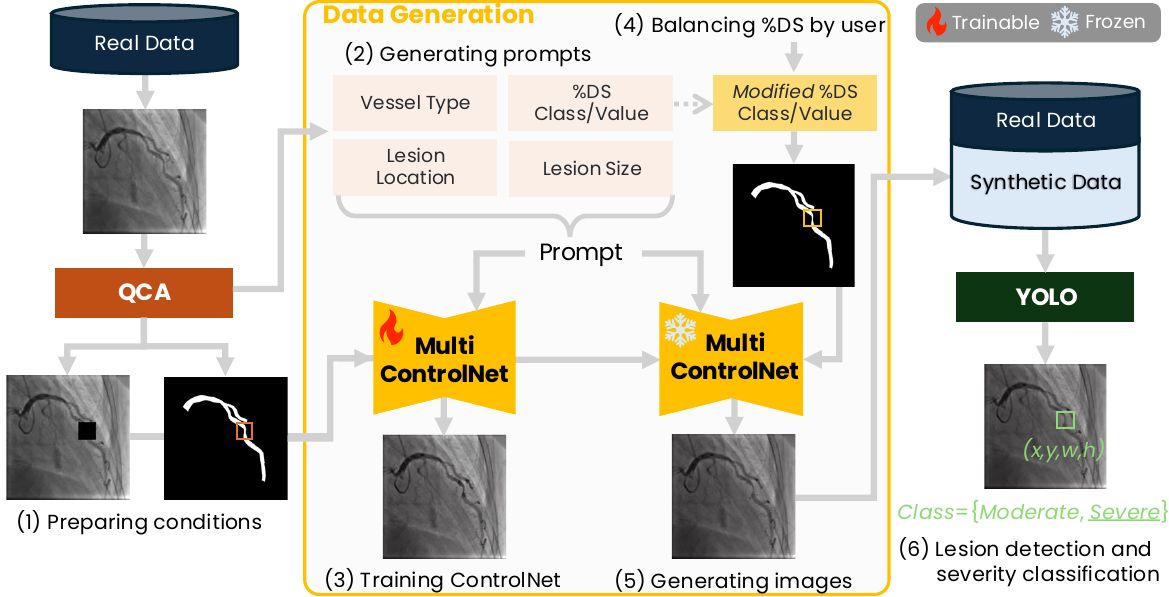}
    \caption{
        \textbf{Overall pipeline of proposed method.} (1,2) We prepare conditioning images and generate prompts from metadata. (3) ControlNet is trained to reconstruct images and (4,5) inferred to generate new images with modified segmentation maps. (6) YOLO model captures lesions and predicts their severity.
    }
    \label{fig:architecture}
\end{figure}

\section{{Proposed Methods}}
Accurate detection of coronary lesions and measurement of \%DS using quantitative coronary analysis (QCA) tools remain both labor-intensive and susceptible to inter-operator variability.
To mitigate these challenges, we develop a one-stage pipeline that performs lesion detection and severity classification--categorized into 50-70\%DS and $\ge$70\%DS based on the clinical severity \cite{writing20222021}--while  augmenting training data with synthetic data through a controllable data generation scheme.

We first apply a conventional QCA tool to each real angiogram to extract lesion location and severity. A user-adjusted severity level is then utilized as a conditional input for a diffusion-based generative model, which synthesizes angiographic images reflecting various degrees of stenosis while preserving the vascular structure. This approach effectively expands the dataset, balances class distributions, and enhances detection model, particularly in light of limited labeled angiographic datasets. Fig.~\ref{fig:architecture} illustrates the overall data augmentation pipeline, including generative model training and inference.

\subsection{Vessel Segmentation and Severity Designation}
Before generating a realistic angiogram image with a specified stenosis level (\%DS), 
 conventional QCA is applied to the original image to extract vessel contours, as shown in Fig.~\ref{fig:qualitative} (g).
 To meet the specific \%DS value, we compute \%DS using the reference vessel diameter ($D_{\text{ref}}$), which is estimated by interpolating the vessel contour from adjacent non-diseased vessel segments, while the minimum lumen diameter (MLD) is identified at the site of maximum stenosis within the lesion~\cite{beatt1988change,brown1977quantitative}. \%DS is then computed as  
$ \,\bigl(1-(\mathrm{MLD} / D_{\mathrm{ref}}\bigr)) \times 100$.

Our algorithm is designed to move two control points at the MLD location by shifting them in a direction orthogonal to the vessel direction such that it matches with the desired \%DS. 
After redrawing the vessel contours, these modifications propagate smoothly to nearby points to ensure a natural transition without abrupt changes. 
The resulting contours are converted into segmentation masks, and they are encoded into a compact latent space and integrated into our conditional generation framework.

\subsection{Data Augmentation with Multi-Control Diffusion Model}
To generate angiograms with lesions that exhibit varying stenosis severity, we leverage ControlNet~\cite{zhang2023controlnet}, an extension of Stable Diffusion~\cite{rombach2022ldm}, which integrates structural guidance to constrain the generation process. 
Unlike previous diffusion-based data augmentation methods~\cite{Cob_Improved_MICCAI2024,fang2024data}, our approach incorporates two conditioning inputs across multiple ControlNets: (1) an original image on which a lesion bounding box is overlaid for inpainting, specifying regions that require synthesis, and (2) a vessel segmentation mask that guides the model to adhere to the vascular structures.
The guidance image $\mathbf{c}$, derived from the segmentation mask and the masked image for inpainting, are encoded into latent representations using a lightweight convolution network, yielding $\mathbf{c}_\text{s}$ and $\mathbf{c}_\text{m}$, respectively.
The network $\epsilon_\theta$, equipped with zero-convolution layers for conditioning, is fine-tuned to predict the noise added to the noisy latent image $\mathbf{z}_t$, incorporating text prompt $\mathbf{c}_\text{t}$ and timesteps $\textbf{t}$, by optimizing the following objective: 
\begin{equation}
    \mathcal{L}=\mathbb{E}_{\mathbf{z}_t, \textbf{t}, \mathbf{c}_\text{t}, \mathbf{c}_\text{s}, \mathbf{c}_\text{m} \sim \mathcal{N}(0,1)}[\| \epsilon - \epsilon_\theta(\mathbf{z}_t, \textbf{t}, \mathbf{c}_\text{t}, \mathbf{c}_\text{s}, \mathbf{c}_\text{m})\|^2].
\end{equation}
The fine-tuned generative model synthesizes the masked structures through two separate denoising processes, ensuring seamless integration of newly generated regions with the original surroundings.

\subsection{Stenosis Detection and Severity Classification}
To enhance detection model performance, we add synthesized images to training dataset, expanding dataset to $N$-times the size of real data to balance the ratio of moderate-to-severe stenosis cases.
Unlike existing multi-step pipelines~\cite{avram2023cathai,kim2024artificial}, which require separate modules for segmenting vessel, stenosis detection and severity prediction, our method streamlines the process into a single-step detection framework leveraging the efficiency of YOLO~\cite{khanam2024yolov11}.
We trained the model on coronary angiogram images, where the detection module generates bounding boxes of lesions with 50\%DS or higher, while the classification module assigns a severity class by computing the probability of belonging to each \%DS category. 

\section{Experiments}

\begin{figure}[ht]
\centering
\begin{minipage}[ht]{0.5\linewidth}
    \centering
    \captionof{table}{\textbf{Data statistics.}}
    \label{tab:stat}
    \resizebox{\linewidth}{!}{
    \begin{tabular}{lrrrrr}
        \toprule
         \multicolumn{1}{c}{} & \multicolumn{3}{c}{\textbf{Internal}} & \multicolumn{1}{c}{\textbf{ARCADE}} \\
            \multicolumn{1}{c}{} & \multicolumn{1}{c}{\textbf{Train}} & \multicolumn{1}{c}{\textbf{Valid}} & \multicolumn{1}{c}{\textbf{Test}} & \multicolumn{1}{c}{\textbf{Valid}}\\
            \midrule
            $\#$Patients & 1402 & 535 & 232 & 200 \\
            $\#$Images & 5156 & 1854 & 884 & 200 \\
            \midrule
            $\#$Lesions & 6030 & 2150 & 1070 & 262 \\
            -- Moderate & 5350 & 1932 & 920 & 236 \\
            -- Severe & 680 & 218 & 150 & 26 \\
        \bottomrule
    \end{tabular}
    }
\end{minipage}
\begin{minipage}[ht]{0.42\linewidth}
    \centering
    \includegraphics[width=\linewidth]{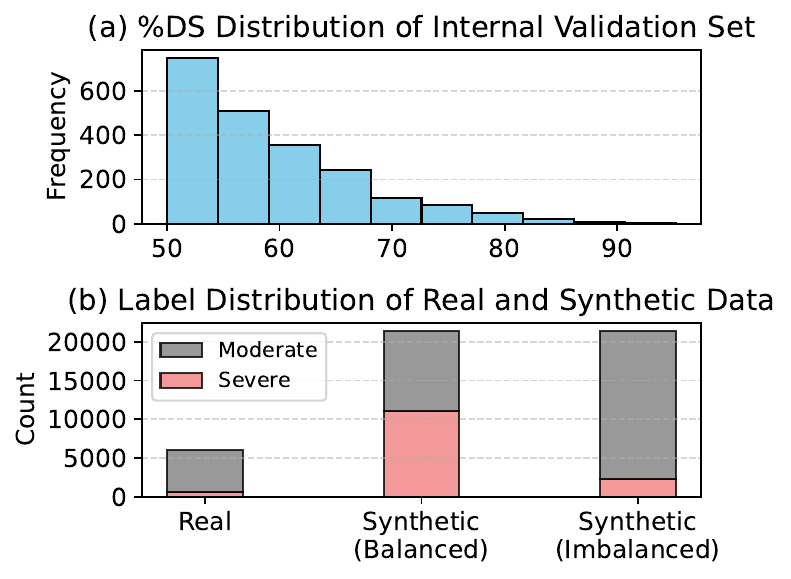}
    \captionof{figure}{
        \textbf{Label distributions.}
    }
    \label{fig:stat}
\end{minipage}
\end{figure}


\subsection{Datasets}
\textbf{Internal dataset} is composed of 7,894 X-ray coronary angiograms (train 5,156, validation 1,854, test 884 images), which is a curated collection of high-resolution angiographic images. We annotated lesions using an automatic QCA tool, with stenosis defined as \%DS greater than 50.
Unlike many existing datasets that rely solely on segmented vessels without specific lesion details, our dataset includes bounding box annotated lesions and corresponding \%DS values, making it uniquely suited for both detection and classification tasks. 
For clinical relevance and adherence to diagnostic standards, QCA annotations are verified by experienced clinicians. 
The \%DS values are categorized into clinically significant ranges (\textit{e}.\textit{g}., <70\%DS, $\geq$70\%DS)~\cite{writing20222021}. Fig.~\ref{fig:stat} illustrates the highly skewed label distributions of our dataset and detailed data statistics are described in Table~\ref{tab:stat}.

To evaluate the generalizability of our method, we also utilize publicly available dataset as an external validation set for our pipeline.
\textbf{ARCADE}~\cite{popov2024arcade} consists of 1,200 coronary angiograms (1,000 train and 200 validation images) from 1,200 different patients with stenosis, defined by $\geq$ 50\%DS. 
One of the limitations of this dataset is that their annotations are segmented into regions, lacking precise regions of interest and reference standards for accurately specifying \%DS values. This ambiguity poses a challenge for evaluating \%DS classification tasks.
To address this, we use the QCA tool to re-annotate the ARCADE validation set and collaborated with our in-house clinicians for refining labels. The labeling process involves identifying lesion locations with corresponding \%DS values, ensuring high-quality and clinically relevant annotations. Furthermore, to foster future research in coronary stenosis detection, we will release the re-annotated dataset for public use at \href{https://github.com/medipixel/DiGDA}{https://github.com/medipixel/DiGDA}.

\subsection{Experimental Setup}
All input images are resized to a standard resolution of $512 \times 512$ to maintain consistency across experiments. 
We use segmentation mask along with bounding boxes as conditioning input to ControlNet~\cite{zhang2023controlnet} component, which is encoded into a $64 \times 64$ latent.
For downstream tasks, we fine-tune a YOLO model on our internal dataset using a learning rate of \(2 \times 10^{-4}\) and a batch size of 32. The model is trained for 50 epochs using the AdamW optimizer to ensure convergence. 
We use the F1-score metric to evaluate the balance between precision and recall for severity classification task and mean Average Precision (mAP50) which evaluates the detection accuracy for lesion detection.
We report the average scores over three runs for all downstream experiments.

\begin{table*}[tp]
    \centering
    \caption{\textbf{Performance on different ratios of real to synthetic data.}
    The real-to-synthetic ratio is expressed as $\times N$, where $N$ denotes the synthetic data scaling factor, 
    while the number of targeted lesions per class is balanced. The mAP50 metrics are also reported for each stenosis severity classes, respectively: Moderate (M, $<70$\%DS) and Severe (S, $\geq70$\%DS).}
    \label{tab:model_performance}
    \begin{tabular}{ccccccccccc}
        \toprule
         \multicolumn{1}{c}{\textbf{Synthetic}} & \multicolumn{4}{c}{\textbf{Internal dataset}} & \multicolumn{4}{c}{\textbf{ARCADE}} \\
        \textbf{data ratio} & \textbf{F1} & \textbf{mAP50} & \textbf{(M} & \textbf{S)} & \textbf{F1} & \textbf{mAP50} & \textbf{(M} & \textbf{S)} \\
            \midrule
            \textbf{$\times$0} & 0.650 & 0.688 & 0.731 & 0.646 & 0.500 & 0.464 & 0.436 & 0.492 \\
            \midrule
            \textbf{$\times$1} & 0.656 & 0.699 & 0.749 & 0.649 & 0.519 & 0.484 & 0.440 & 0.527 \\
            \textbf{$\times$2} & 0.658 & 0.710 & 0.767 & 0.654 & \textbf{0.524} & \textbf{0.501} & \textbf{0.448} & \textbf{0.555} \\
            \textbf{$\times$4} & \textbf{0.670} & \textbf{0.717} & \textbf{0.771} & \textbf{0.663} & 0.513 & 0.492 & 0.438 & 0.546 \\
        \bottomrule
    \end{tabular}
\end{table*}


\subsection{Quantitative Evaluation}

Table~\ref{tab:model_performance} summarizes the performance of our method (\OURS) compared to a baseline detector trained without data augmentation using synthetic images. 
\OURS demonstrates superior results on both internal and external dataset, achieving higher precision, recall, and mAP scores, for the increased size of balanced set with $\times1$, $\times2$, and $\times4$ data ratio. For the synthetic data setting $\times4$, mAP50 increased from 0.688 to 0.717, resulting in a significant lesion detection performance gain.
External validations have shown similar trends with consistently improved performance compared to the baseline model trained with $\times0$ setting.
For each setting ($\times N$), synthetic data equivalent to N times the amount of real data is first generated, followed by a removal of a portion of the generated data to ensure that the number of samples for each label remains equal across settings. This highlights the effectiveness of incorporating synthetic data generated using our approach.

\begin{figure}[tp]
    \centering
    \includegraphics[width=\linewidth]{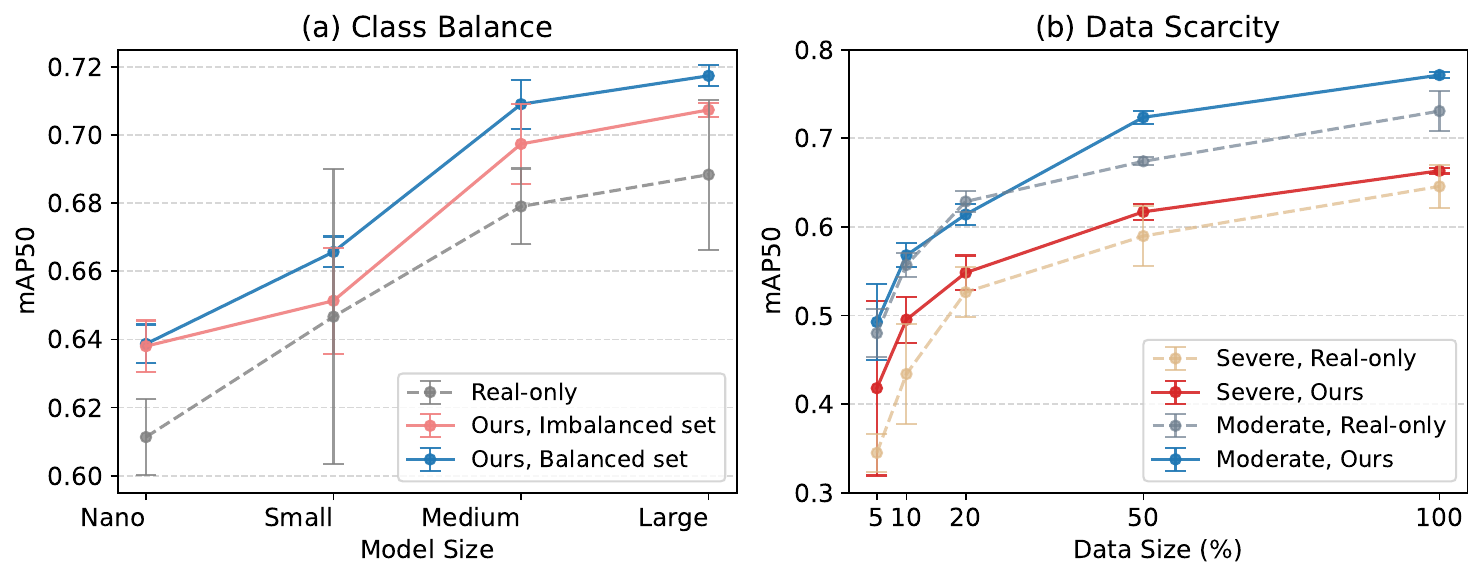}
    \captionof{figure}{
        \textbf{Performance comparisons on lesion detection across data compositions and dataset sizes.}
        (a) examines the effect of class balance: the balanced set has equal target lesions per class, while the imbalanced set maintains the original distribution with the same amount of the total data.
        (b) compares model performance across different subset sizes.
    }
    \label{fig:ablation_subset}
\end{figure}

Furthermore, we analyze the effect of dataset composition by comparing a class-balanced dataset to an imbalanced dataset for detection task, as shown in Fig.~\ref{fig:ablation_subset} (a). Unlike previous experiments that maintained class balance, this ablation introduces class imbalance, mirroring the label distribution of real data where class frequencies are naturally skewed. 
\OURS, when trained on a balanced dataset, consistently achieves superior detection performance in different model sizes, outperforming models trained on imbalanced data. 
\OURS, based on YOLO large model, trained with a balanced dataset augmented by $\times 4$ yields a mAP50 gain with large gap compared to the model trained solely on real data, and a mAP50 increase from 0.707 to 0.717 compared to the imbalanced dataset.

As illustrated in Fig.~\ref{fig:ablation_subset} (b), the proposed method improves performance under different subset size settings. Our approach consistently maintains higher mAP50, with notable improvements in severe stenosis class, which has been extensively enriched through data augmentation. Note that both the generative model and the detection model are trained on a subset.

\subsection{Qualitative Results}
We conduct a qualitative analysis to visualize the capability of our method as illustrated Fig.~\ref{fig:qualitative}. 
Our approach successfully generates realistic vessel narrowing patterns corresponding to various levels of \%DS, ensuring anatomical plausibility and alignment with clinical expectations. 
Compared to the original images with \%DS values computed by conventional QCA, our generated images exhibit higher \%DS which represents the severe stenosis category, underrepresented in the dataset.
The results show that our method accurately captures vessel constriction patterns while maintaining structural consistency across different vessel types. By using narrower vessel masks modified from original ones by user-defined \%DS, our approach enables precise severity modulation overcoming the limitations of existing datasets.
Furthermore, our inpainting strategy ensures that background regions, except for the areas within the red-outlined bounding boxes, remain unchanged, preserving the original angiographic context. 
This targeted modification ensures that an augmented dataset without incurring additional labeling cost for refinements.

\section{Conclusion}
In this study, we propose a novel approach to improve lesion detection and severity classification, integrating synthetic data generated via a diffusion model to vessel narrowing simulations into the training process. 
User-guided modifications to vessel masks enable the simulation of various severity levels that are underrepresented in the real-world datasets. 
By leveraging both internal and external datasets, we demonstrate the efficacy of synthetic data in improving model performance, particularly in scenarios where real-world data is scarce or class-imbalanced. Our method achieves significant improvements in detection and classification, as confirmed in both quantitative and qualitative evaluations.
The inclusion of synthetic data allows the model to learn various stenosis patterns while faithfully preserving the vascular structural integrity. 
Overall, this work demonstrates the potential of integrating synthetic data into lesion detection pipelines, paving the way for enhanced diagnostic tools in medical imaging.

\textbf{Acknowledgment.}
The internal dataset was approved by the Institutional Review Boards of the participating hospitals, with informed consent waived, and collected in accordance with the Declaration of Helsinki. 
A retrospective data collection was performed on consecutive patients who underwent CAG at Sapporo Cardio Vascular Clinic (Sapporo, Japan; 2018–2023), Veterans Health Service Medical Center (Seoul, Republic of Korea; 2013–2023), and Soonchunhyang University Cheonan Hospital (Cheonan, Republic of Korea; before 2024).
This work was supported by the Korea Medical Device Development Fund grant funded by the Korean government (the Ministry of Science and ICT, the Ministry of Trade, Industry and Energy, the Ministry of Health and Welfare, and the Ministry of Food and Drug Safety)
(Project Number: RS-2023-00223121).

\bibliographystyle{splncs04}
\bibliography{main}

%




\end{document}